\pdfoutput=1
\documentclass[preprint,preprintnumbers,amsmath,amssymb,superscriptaddress]{revtex4}
 
\newcommand{\be}{\begin{equation}}
\newcommand{\ee}{\end{equation}}
\newcommand{\ba}{\begin{eqnarray}}
\newcommand{\ea}{\end{eqnarray}}
\newcommand{\bd}{\begin{displaymath}}
\newcommand{\ed}{\end{displaymath}}

\newcommand{\commentout}[1]{{}}

\def\thalf{{\textstyle{\frac{1}{2}}}}
\def\oneth{{\textstyle{\frac{1}{3}}}}

\usepackage{amssymb}
\usepackage{color}
\usepackage{slashed}
\usepackage{amsmath}
\usepackage{graphicx}
\usepackage{epstopdf}
\usepackage{dcolumn}
\usepackage{bm}
\usepackage{hyperref}

\begin{document}

\title{{\bf Causal Baryon Diffusion and Colored Noise}}
\author{{J. I. Kapusta  and C. Young} \vspace*{0.1in}\\
{\it School of Physics and Astronomy, University of Minnesota}\\
 {\it Minneapolis, Minnesota 55455, USA}}


\date{\today}

\begin{abstract}

We construct a model of baryon diffusion which has the desired properties of causality and analyticity.  The model also has the desired property of colored noise, meaning that the noise correlation function is not a Dirac delta function in space and time; rather, it depends on multiple time and length constants.  The model can readily be incorporated in 3+1 dimensional second order viscous hydro-dynamical models of heavy ion collisions, which is particularly important at beam energies where the baryon density is large.

\end{abstract}

\maketitle

\section{Introduction}

The fluctuation-dissipation theorem requires all dynamical systems in or near thermal equilibrium to experience noise.   Hydrodynamical fluctuations in non-relativistic viscous fluids have been understood for some time \cite{Landau:1980st}.  Fluctuations can be especially important in small systems, such as in ordinary liquids whose dimensions are on the order of nanometers \cite{nanojet,nanobridge}.   

Hydrodynamics is a state of the art tool to describe high energy heavy ion collisions where relativity is crucial.  In these collisions the dimensions are on the order of 5 to 15 fm, and so hydrodynamical fluctuations ought to be important.  Noise in relativistic hydrodynamics was worked out in ref. \cite{Kapusta:2011gt} and applied to the ubiquitous Bjorken model.  It was found that noise contributes to two-particle correlations, resulting in a ridge-like structure in rapidity.  These and analogous two-particle correlations in azimuth might be a way to obtain an independent measurement of the shear and bulk viscosites.  It might also be a way to infer the thermal conductivity and the existence of a critical point in the QCD phase diagram \cite{JoeJuan}.

Fluctuating hydrodynamics encounters a singularity which is not encountered in noiseless hydrodynamics \cite{Young:2013fka}. This singularity can be easily understood: the autocorrelation function of the noise in the energy-momentum tensor, $\left \langle \Xi^{\mu \nu}({\bf x}, t)\Xi^{\rho \sigma}({\bf x}^\prime, t^\prime) \right\rangle$, is proportional to a four-dimensional Dirac delta function. This is white noise in frequency and momentum space.  It means that the integrated noise in a cell of space-time volume $\Delta V \Delta t$ has a root-mean-square proportional to $\sqrt{\Delta V \Delta t}$.  As a consequence, the average value of the noise in this cell diverges as the discretization in space and time is made small, being proportional to $1/\sqrt{\Delta V \Delta t}$. Such a divergence leads to large gradients which call the gradient expansion at the heart of hydrodynamics into question. Specifically, simulations of heavy-ion collisions which exist at the lower limit in size of systems describable with hydrodynamics are severely limited in resolution when thermal noise is included in the most straightforward way. Treating hydrodynamic fluctuations as a perturbation solves the problem of this divergence on a practical level, and indeed that is what is done analytically in \cite{Kapusta:2011gt} and \cite{JoeJuan}.  However, the question of what this divergence implies for a maximum resolving power of hydrodynamics remains.

Reference \cite{Young:2013fka} also determined the thermal noise in second-order viscous hydrodynamics in the Israel-Stewart formalism \cite{Israel}; a similar approach is also taken in \cite{Murase:2013tma}. The autocorrelation function of the noise is smoothened in time by the relaxation time $\tau_\pi$, a second-order transport coefficient, so that the autocorrelation function is proportional to 
$\delta^3({\bf x}-{\bf x}^\prime) \, {\rm e}^{-|t-t^\prime|/\tau_\pi}$.  This established the shape of thermal noise in \textsc{music} \cite{Schenke:2010}, a numerical hydrodynamical code which implements the Israel-Stewart formalism.  Unfortunately there is still a three-dimensional spatial delta function so the problem remains. 

The goal of this paper is to focus on the general problem of baryon number diffusion, fluctuation, and noise with potential applications to heavy ion collisions.  This is very relevant to past and future experiments of beam energy scans (BES) at the Relativistic Heavy Ion Collider (RHIC), at the Facility for Antiproton and Ion Research (FAIR), at the SPS Heavy Ion and Neutrino Experiment (SHINE), and at the Nuclotron-based Ion Collider Facility (NICA).   Our work is applicable to any conserved current and can readily be generalized to include more than one conserved charge, such as electric charge and isospin. 

Diffusion of a conserved charge, such as baryon number, is distinct from but closely related to heat diffusion.  As was known already by Maxwell in the 19th century, the diffusion equation for heat propagates a signal at infinite speed, which is unphysical even apart from relativity.  Maxwell did not consider this a problem of practical concern for experiments of the day; see ref. \cite{Joseph} for a history of this issue.  

The outline of our paper is as follows.  In sect. II we consider the baryon diffusion equation and its generalization in successive powers of derivatives in space and time.  These lead to the Cattaneo equation \cite{Cattaneo} and to the Gurtin-Pipkin equation \cite{GP}, both of which were proposed as models of heat conduction, not baryon diffusion.  We show how these relate to the baryon density response function, to baryon density fluctuations, and to baryon noise.  In sects. III, IV and V we show how these general considerations apply to the conventional diffusion equation, to the Cattaneo equation, and to the Gurtin-Pipkin equation, respectively.

We find that both the Cattaneo and Gurtin-Pipkin approaches lead to finite speed of propagation for the baryon density autocorrelation function.  However, only the Gurtin-Pipkin approach leads to smeared out correlations in space and time for the noise, and thus is preferable for modeling high energy nuclear collisions.  Our conclusions are presented in sect. VI.   

\section{General Considerations}

In this section we outline three approaches to the problem of baryon diffusion with successively increasing number of space and time derivatives.  Detailed calculations of the response function, fluctuations, and noise for each approach are considered in subsequent sections.

The Landau-Lifshitz approach is the most commonly used one for high energy heavy ion collisions.  In this approach $u^{\mu}$ is defined to be the velocity of energy transport, whereas in the Eckart approach it is defined to be the velocity of baryon transport.  The Landau-Lifshitz approach is favored because the baryon density is small in comparison to the energy and entropy densities, sometimes making the definition of flow velocity in the Eckart approach problematic.  The baryon current takes the form
\be
J^{\mu} = n u^{\mu} + \Delta J^{\mu}
\ee
where $n$ is the proper local baryon density and $\Delta J^{\mu}$ is the dissipative part.  This modification to the current must satisfy $u_{\mu} \Delta J^{\mu} = 0$ in order that $n$ represent the proper baryon density. 

In first order viscous fluid dynamics $\Delta J^{\mu}$ takes the form
\be
\Delta J^{\mu} = \sigma T \Delta^{\mu} \left(\beta \mu \right) \, ,
\ee
where $\beta = 1/T$, $\mu$ is the chemical potential, $\sigma$ is the baryon conductivity and
\be
\Delta_{\mu} = \partial_{\mu} - u_{\mu} \left( u \cdot \partial \right)
\ee
is a derivative normal to $u^{\mu}$.  For baryon diffusion in a system with no energy flow one obtains the usual diffusion equation
\be
\left[ \frac{\partial}{\partial t} - D\nabla^2  \right] n = 0
\ee  
where the diffusion constant and baryon conductivity are related by $\sigma = D (\partial n/\partial \mu)$.  As is well known, the diffusion equation results in instantaneous transport and is not suitable for numerical hydrodynamic simulations of high energy heavy ion collisions.

It has been suggested to replace the usual diffusion equation by the second order hyperbolic equation
\be
\left[ \frac{\partial}{\partial t} - D\nabla^2 + \tau \frac{\partial^2}{\partial t^2} \right] n = 0
\ee
which involves a characteristic time scale $\tau$.  This equation is also recognized as the telegraph equation.  It's application to heat transport is generally attributed to Cattaneo (although Maxwell did consider the second order time derivative but dropped it on the grounds that it is irrelevant in practice).  It arises from a modification to the dissipative part of the current in the form
\be
\Delta J^\mu = D \Delta^{\mu} \frac{1}{1+\tau(u \cdot \partial)}\,  n \equiv D \Delta^{\mu} \sum_{l=0}^{\infty} \left[ - \tau (u \cdot \partial) \right]^l n
\ee
which is of infinite order in time derivatives (in the local rest frame).  At high frequency, waves travel with speed $v_0 = \sqrt{D/\tau}$.  Although this description leads to a baryon density response function that has the required features of causality and analyticity, it does not lead to a noise correlator which has a finite correlation length as we shall see later.

Going to the third order in time and space derivatives results in an equation first applied to the problem of heat conduction by Gurtin and Pipkin.
\be
\left[ \frac{\partial}{\partial t} - D\nabla^2 + \tau_1 \frac{\partial^2}{\partial t^2} + \tau_2^2 \frac{\partial^3}{\partial t^3}
- \tau_3' D\frac{\partial}{\partial t}\nabla^2  \right] n = 0
\ee
(The reason for the prime will become apparent.)  This equation is also hyperbolic. High frequency waves travel with speed $v_0 = \sqrt{\tau_3' D/\tau_2^2}$.   This equation follows from the dissipative current
\be
\Delta J^\mu = D \Delta^{\mu} \frac{1 + \tau_4(u \cdot \partial)}{1+\tau_1(u \cdot \partial) + \tau_2^2(u \cdot \partial)^2
 + \tau_3 D \Delta^2} \, n
\ee
where again the differential operator in the denominator is to be understood as its Taylor series expansion.  Note that there are four time constants in the current as $\tau_3' = \tau_3 + \tau_4$.  Obviously, setting $\tau_2 = \tau_3 = \tau_4 = 0$ results in the Cattaneo equation, and further setting $\tau_1 = 0$ results in the ordinary diffusion equation.  Setting only $\tau_2 = 0$ results in a differential equation of the Jeffrey's type, but it is not hyperbolic and will not be considered here.
  
Suppose that the chemical potential is varied by an amount $\delta \mu$ by some external source.  This results in a change in energy $\delta H = \int d^3x\, n \, \delta\mu$.  The current then satisfies the equation
\be
\partial_{\mu} J^{\mu} = \left(\frac{\partial n}{\partial \mu} \right) \frac{\partial \delta\mu}{\partial t}
\ee
Let a thermodynamic quantity in frequency and wavenumber space be denoted with a tilde, and let $\delta n$ denote the deviation from the uniform background density.  Then, in Fourier space
\be
\left[ -i\omega + \frac{Dk^2(1-i\tau_4 \omega)}{1-i\tau_1\omega - \tau_2^2 \omega^2 +\tau_3 D k^2}\right]\delta n = 
-i   \left(\frac{\partial n}{\partial \mu} \right) \omega \delta \mu
\ee
The response function for $\delta\tilde{n}$ comes from the ratio of the terms above, namely
\be
G_R(\omega, {\bf k}) = \left(\frac{\partial n}{\partial \mu} \right) \frac{\omega}{A(\omega, {\bf k})}
\ee
where 
\be
A(\omega, {\bf k}) \equiv \omega + \frac{iDk^2(1-i\tau_4 \omega)}{1-i\tau_1\omega - \tau_2^2 \omega^2+ \tau_3 D k^2}{\rm .}
\label{A}
\ee
The autocorrelation function is 
\be
\left\langle \delta n \delta n(\omega, {\bf k}) \right \rangle = -\frac{2T}{\omega}{\rm Im}\left\{ G_R \right\} 
= 2 T \left(\frac{\partial n}{\partial \mu} \right) \frac{{\rm Im}\left\{ A \right\}}{|A|^2}
= i T \left(\frac{\partial n}{\partial \mu} \right) \left( \frac{1}{A} - \frac{1}{A^*} \right)
\ee
Note that both the response function and the autocorrelation function for the baryon density will generally have poles at the zeroes of $A(\omega, {\bf k})$ and $A^*(\omega, {\bf k})$, depending on the values of the parameters.

Now let us consider the autocorrelation function for the noise.  From current conservation $\partial_{\mu}J^{\mu}_{\rm total} = 0$, where $J^{\mu}_{\rm total} = J^{\mu} + I^{\mu}$, $J^{\mu} = n u^{\mu} + \Delta J^{\mu}$, $\Delta J^{\mu}$ is the dissipative part, and $I^{\mu}$ is the noisy part, we have
\be
\langle \partial_{\mu}J^{\mu}({\bf x},t) \partial_{\nu}J^{\nu}({\bf 0},0) \rangle
= \langle \partial_{\mu}I^{\mu}({\bf x},t) \partial_{\nu}I^{\nu}({\bf 0},0) \rangle
\ee
We work in the rest from of the fluid, $u^0 =1$, $u^i = 0$.  The Cartesian components of the noise current are independent so that
\be
\left\langle I^i({\bf x},t)I^j(0, {\bf 0}) \right\rangle = \oneth \left\langle I^l({\bf x},t) I^l(0, {\bf 0}) \right\rangle \delta_{ij}
\ee
After Fourier transforming we have
\be
\oneth k^2 \langle I^l I^l ({\bf k},\omega) \rangle =  
A(\omega, {\bf k}) A^*(\omega, {\bf k}) \langle \delta n \delta n({\bf k},\omega \rangle =
- i T \left(\frac{\partial n}{\partial \mu} \right) \left( A - A^* \right)
\label{gen_noise_cor}
\ee
The noise correlator has the same singularities as $A - A^*$, whereas the baryon autocorrelation function has singularities at the zeroes of $A$ and $A^*$.

\section{Ordinary Diffusion Equation}

For the ordinary diffusion equation
\be
A = \omega + iDk^2
\ee
The response function
\be
G_R = \left(\frac{\partial n}{\partial \mu} \right) \frac{\omega}{\omega + iDk^2}
\ee
has a simple pole in the lower half plane and is analytic in the upper half plane.  Therefore, it is causal in the sense that if a disturbance occurs at $t=0$, there is no response at negative times.  The baryon autocorrelation function in time and wavenumber space is
\be
\left\langle \delta n \delta n(t, {\bf k}) \right \rangle = T \left(\frac{\partial n}{\partial \mu} \right) 
{\rm e}^{-D k^2 t}
\label{Dkt}
\ee
In time and coordinate space it is
\be
\left\langle \delta n \delta n(t, {\bf x}) \right \rangle = T \left(\frac{\partial n}{\partial \mu} \right) 
\left(\frac{1}{4 \pi D t}\right)^{3/2} {\rm e}^{-r^2/4 D t}
\ee
Thus, although it is causal in the sense mentioned before, baryon diffusion happens with infinite speed of propagation.

In frequency and wavenumber space the noise correlator is just a constant.
\be
\oneth \langle I^l I^l ({\bf k},\omega) \rangle = 2 \sigma T
\ee
In time and coordinate space it is
\be
\langle I^i I^j (t, {\bf x}) \rangle = 2 \sigma T  \delta ({\bf x}) \, \delta(t) \, \delta_{ij}
\label{Dnoise}
\ee
This is identical to the result obtained in \cite{Kapusta:2011gt}.

\section{Cattaneo Equation}

To obtain the Cattaneo equation means setting $\tau_2 = \tau_3 = \tau_4 = 0$.  Hereafter in this section we shall relabel $\tau_1$ as $\tau_D$.

Now
\be
A = \omega + \frac{iDk^2}{1-i\tau_D\omega}
\ee
and the response function can be written as
\be
G_R =   \left(\frac{\partial n}{\partial \mu} \right) \frac{\omega (\omega + i/\tau_D)}{ (\omega - \omega_+) (\omega - \omega_-)}
\ee
Here 
\ba
\omega_{\pm} &=& - \frac{i}{2\tau_D} \pm \frac{i \delta_k}{2\tau_D} \nonumber \\
\delta_k &=& \sqrt{1-4\tau_D D k^2} \le 1
\ea
when $k \le k_c \equiv 1/2\sqrt{\tau_D D}$, and
\ba
\omega_{\pm} &=& - \frac{i}{2\tau_D} \pm \frac{\epsilon_k}{2\tau_D} \nonumber \\
\epsilon_k &=& \sqrt{4\tau_D D k^2 - 1} \ge 0
\ea
when $k > k_c$.  The fact that $G_R$ is analytic in the upper half-plane is a consequence of analyticity.

The group velocity is only defined when $k > k_c$.  It is
\be
v_g(k) = \frac{1}{2 \tau_D} \frac{\partial \epsilon_k}{\partial k} = \frac{2Dk}{\sqrt{4\tau_D D k^2 - 1}}
\ee
which goes to $v_0 \equiv \sqrt{D/\tau_D}$ as $k \rightarrow \infty$.  This means that $\tau_D > D$ if the group velocity is not to exceed the speed of light in the ultraviolet.  The fact that the group velocity exceeds the speed of light for a finite range of $k$ is not a problem.  Recall that the group velocity arises from a Taylor series expansion around the centroid of a wave packet.  The interpretation of $d\omega/dk$ as the propagation of a signal is predicated on the assumption that  $d^2 \omega/dk^2$ is small.  That assumption is violated in the vicinity of $k_c$ where $v_g(k)>1$.  See ref. \cite{Brillouin}.

\subsection{Baryon correlation function}

The equal-time autocorrelation function for density perturbations is
\ba
\langle \delta n \delta n(0,{\bf x}) \rangle &=& - 2 T \left(\frac{\partial n}{\partial \mu}\right) \int \frac{d^3k}{(2\pi)^3}
{\rm e}^{i {\bf k}\cdot{\bf x}} \; {\rm Im} \int \frac{d\omega}{2\pi} \frac{\omega + i/\tau_D}{(\omega - \omega_+)
(\omega - \omega_-)} \nonumber \\ 
&=& 2 T \left(\frac{\partial n}{\partial \mu}\right) \delta ({\bf x})
\ea
which is independent of $D$ and $\tau_D$.  This is just the usual thermal average, as it must be.

Now let's consider the autocorrelation function for positive time $t > 0$.  The Fourier transform is
\be
\int_{-\infty}^{\infty} \frac{d\omega}{2\pi} {\rm e}^{-i\omega t} \langle \delta n \delta n(\omega,{\bf k}) \rangle =  T \left(\frac{\partial n}{\partial \mu}\right) 
{\rm e}^{-t/2\tau_D} S(k,t)
\label{FTC1}
\ee
where
\be
S(k,t) = 
\left\{ \begin{array}{ll}
{\displaystyle \cosh\left(\frac{\delta_k t}{2\tau_D}\right) +  \frac{1}{\delta_k} \sinh\left(\frac{\delta_k t}{2\tau_D}\right)} & \; {\rm if} \; k \le k_c \\
{\displaystyle  \cos\left(\frac{\epsilon_k t}{2\tau_D}\right) +  \frac{1}{\epsilon_k} \sin\left(\frac{\epsilon_k t}{2\tau_D}\right)} & \; {\rm if} \; k \ge k_c
\end{array} \right.
\label{FTC2}
\ee
which is an even, analytic function of $k$.  Fourier transforming in space gives
\be
\int \frac{d^3k d\omega}{(2\pi)^4} {\rm e}^{i({\bf k}\cdot{\bf x}-\omega t)} \langle \delta n \delta n(\omega,{\bf k}) \rangle
 =  \frac{T}{2 \pi^2 r} \left(\frac{\partial n}{\partial \mu}\right) {\rm e}^{-t/2\tau_D}
\int_0^{\infty} dk \, k \sin(kr) S(k,t)
\label{FTG}
\ee
Due to the symmetry of the integrand in eq. (\ref{FTG}) the integral can also be written as
\bd
\frac{1}{2i} \int_{-\infty}^{\infty} dk \, k {\rm e}^{ikr} S(k,t)
\ed
For large $k$ the integrand behaves as ${\rm e}^{ik(r \pm v_0t)}$ so if $r > v_0t$ one can add a semi-circular contour in the upper half-plane and apply the residue theorem to show that the autocorrelator vanishes.  This is a manifestation of causality.

To display some numerical results it is useful to use the dimensionless variables $\hat{t} = t/2\tau_D$, $\hat{r} = r/2\sqrt{\tau_D D}$, and $\hat{k} = 2\sqrt{\tau_D D} k$.  Removing all the unimportant prefactors
\be
\langle \delta n \delta n(t,{\bf x}) \rangle = \frac{1}{16 \pi^2} \left(\frac{\partial n}{\partial \mu}\right) \frac{T}{\sqrt{\tau_D^3 D^3}} \, f(\hat{r},\hat{t})
\ee
where
\be
f(\hat{r},\hat{t}) = \frac{{\rm e}^{-\hat{t}}}{\hat{r}} \int_0^{\infty} d\hat{k} \, \hat{k} \sin(\hat{k}\hat{r}) S(\hat{k},\hat{t})
\label{Gbar}
\ee
This illustrates that the characteristic length scale is $\sqrt{\tau_D D}$ while the charactersitic time scale is $\tau_D$; unsurprisingly, they are related by $\sqrt{\tau_D D} = v_0 \tau_D$.    

\begin{figure}
  \centering
  \includegraphics{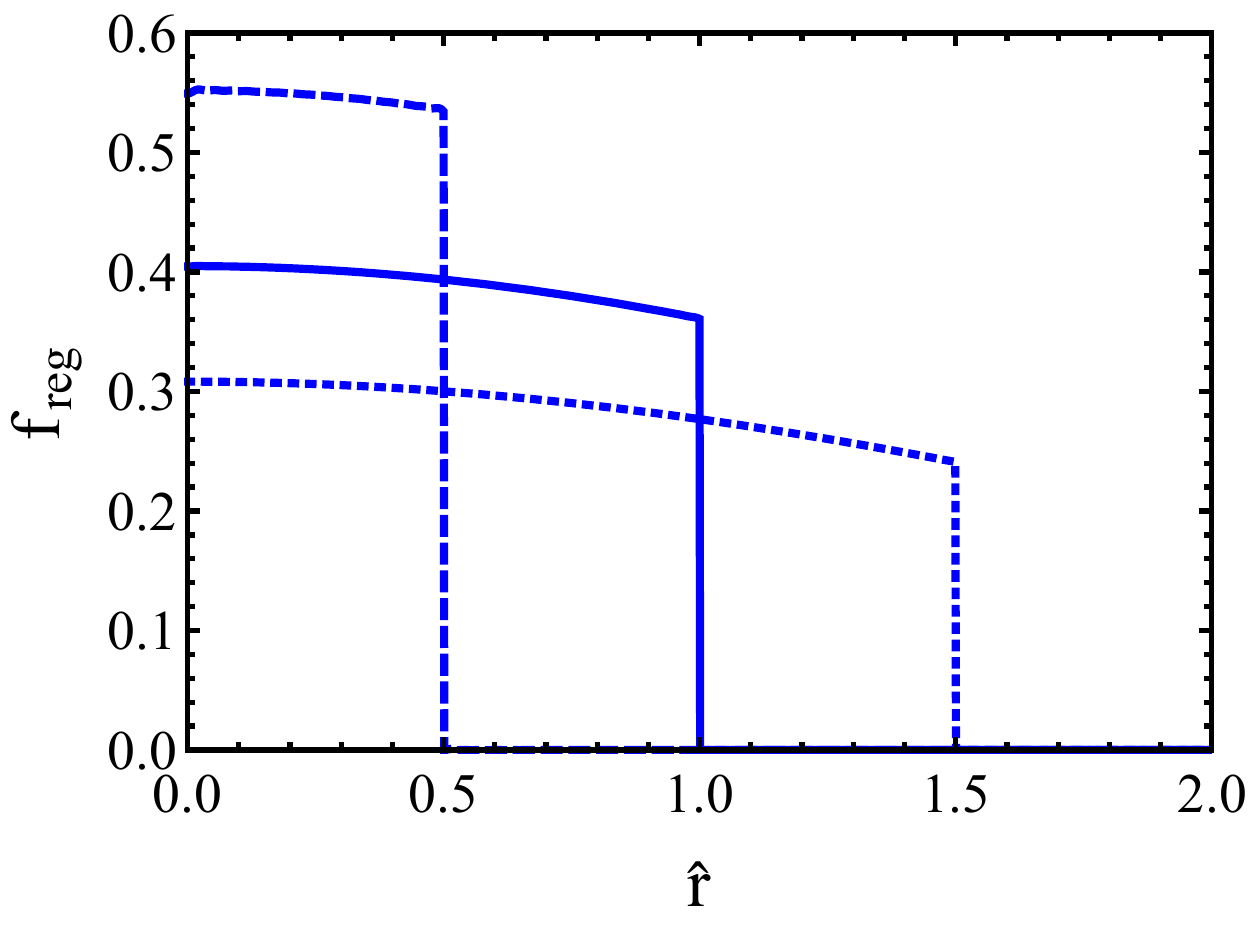}
  \caption{The regular part of the dimensionless baryon correlator $f_{\rm reg}(\hat{r},\hat{t})$ for the Cattaneo equation.
  The correlator is shown for $\hat{t}=0.5$ (dashed line),  $\hat{t}=1$ (solid line), and  $\hat{t}=1.5$ (dotted line).}
  \label{fig:nn_JK}
\end{figure}

Hidden within the integral are singularities at the point $\hat{r}=\hat{t}$.  These may readily be extracted by expanding the integrand up to and including the order $1/\hat{k}$ for large $\hat{k}$.  This is
\be
\hat{k} S(\hat{k},\hat{t}) \rightarrow \left[1 + \frac{\hat{t}}{2} \right] \sin(\hat{k}\hat{t})
+ \left[ \hat{k} - \frac{(4+\hat{t})\hat{t}}{8\hat{k}} \right] \cos(\hat{k}\hat{t})
\label{G at large k}
\ee
These terms contribute to $f$ as
\bd
\frac{\pi}{2} \frac{{\rm e}^{-\hat{t}}}{\hat{r}} \left[ \left(1 + \frac{\hat{t}}{2}\right) \delta (\hat{r}-\hat{t}) - \delta' (\hat{r}-\hat{t})
 -  \frac{(4+\hat{t})\hat{t}}{8} \theta (\hat{r}-\hat{t})  \right]
\ed
The terms involving the Dirac delta function and its first derivative comprise the singular part $f_{\rm sing}$.  The terms in (\ref{G at large k}) are subtracted from the integrand of (\ref{Gbar}) to yield an integral which is easily computed numerically; when the step function is added they together comprise the regular part.  (This procedure avoids the Gibbs phenomenon when representing the step function with a Fourier series.)  Some sample curves are shown in Figure \ref{fig:nn_JK}.  Indeed $f(\hat{r}>\hat{t}) = 0$.  The appearance of the Dirac delta function and its derivative, followed by a diffusion wake, is very similar to what was found for a different response function in \cite{Kapusta:2011gt}.

The conventional diffusion result is recovered in the limit that $r$ is held fixed and $t \gg \tau_D$.  Looking back at eqs. (\ref{FTC1}) and (\ref{FTC2}) it is apparent that in this limit the autocorrelation function is dominated by those values of $k$ which maximize $\delta_k$, namely, $k \ll k_c$.  Then
\be
S(k,t) \rightarrow {\rm e}^{t/2\tau_D} \, {\rm e}^{-D k^2 t}
\ee
which, together with eq. (\ref{FTC1}), leads to the convential result of eq. (\ref{Dkt}).

\subsection{Noise}

In frequency and wavenumber space the noise correlator is
\be
\oneth \langle I^l I^l ({\bf k},\omega) \rangle = \frac{2 \sigma T}{1+(\tau_D \omega)^2}
\ee
In time and coordinate space it is
\be
\langle I^i I^j (t, {\bf x}) \rangle = \frac{\sigma T}{\tau_D} \delta ({\bf x}) \, {\rm e}^{-|t|/\tau_D} \, \delta_{ij}
\label{Cnoise}
\ee
Thus, noise decays exponentially in time but is still a Dirac delta function in space.  In the limit that $\tau_D \rightarrow 0$, eq. (\ref{Cnoise}) goes to eq. (\ref{Dnoise}).

\section{Gurtin-Pipkin Equation}

Now we keep all four $\tau_i$ nonzero.  We will generally assume that $D$, $\tau_1$, $\tau_2^2$, $\tau_3$, and $\tau_3'$ are all positive.  The function $A$ is given by eq. (\ref{A}).  Giving it a common denominator results in the form
\be
A = \frac{\omega -i\tau_1 \omega^2 - \tau_2^2 \omega^3 + i D k^2 + \tau_3' D k^2 \omega}
{1 - i\tau_1 \omega - \tau_2^2 \omega^2 + \tau_3 D k^2}
\ee
The poles of the response function $G_R$ are given as the solutions of the cubic equation
\be
\omega^3 + i \frac{\tau_1}{\tau_2^2} \, \omega^2 - \left(\frac{1 + \tau_3' D k^2}{\tau_2^2}\right) \omega
 - i \frac{D k^2}{\tau_2^2} = 0
\ee
The solutions may be expressed as follows.
\ba
\omega_1 &=& \frac{\sqrt{3}}{2} \left( R_+ + R_- \right) - \frac{i}{2} \left( R_+ - R_- \right) - \frac{i\tau_1}{3\tau_2^2}
\nonumber \\
\omega_2 &=& -\frac{\sqrt{3}}{2} \left( R_+ + R_- \right) - \frac{i}{2} \left( R_+ - R_- \right) - \frac{i\tau_1}{3\tau_2^2}
\nonumber \\
\omega_3 &=& i \left( R_+ - R_- \right) - \frac{i\tau_1}{3\tau_2^2}
\ea
Here
\be
R_{\pm} = \left[ \frac{1}{2}\left( \sqrt{Q^2 + 4 P^3} \pm Q \right) \right]^{1/3}
\ee
with
\be
P = \frac{1}{3\tau_2^2} \left( 1 - \frac{\tau_1^2}{3 \tau_2^2} + \tau_3' D k^2 \right)
\ee
and 
\be
Q = \frac{\tau_1}{3\tau_2^4} \left( 1 - \frac{2\tau_1^2}{9 \tau_2^2} + \tau_3' D k^2 \right) - \frac{D k^2}{\tau_2^2}
\ee
A more compact way of expressing these is to define dimensionless variables $x \equiv \tau_1/\tau_2$, $y \equiv \tau_3' D k^2$, and $z \equiv \tau_2 D k^2$.  Then
\ba
\omega_1 \tau_2 &=& {\rm w}_0 -i {\rm w}_1 \nonumber \\
\omega_2 \tau_2 &=& -{\rm w}_0 -i {\rm w}_1 \nonumber \\
\omega_3 \tau_2 &=& -i {\rm w}_3
\ea
Here
\ba
{\rm w}_0 &=& {\textstyle{\frac{\sqrt{3}}{2}}} (r_+ + r_-) \nonumber \\
{\rm w}_1 &=& \oneth x + \thalf (r_+ - r_-) \nonumber \\
{\rm w}_3 &=& \oneth x - (r_+ - r_-)
\label{wB}
\ea
with
\be
r_{\pm} = \left[ \thalf \left( \sqrt{q^2 + 4 p^3} \pm q \right)\right]^{1/3}
\ee
\be
p = \oneth \left( 1 - \oneth x^2 + y \right)
\ee
\be
q = \oneth \left( 1 - {\textstyle{\frac{2}{9}}} x^2 + y \right) x -z
\ee
Let us examine the behavior of these solutions as functions of $x$, $y$ and $z$.

When $q^2 + 4 p^3 > 0$ there is one imaginary root and a pair of complex roots.  Clearly the complex roots are $\omega_1$ and $\omega_2$.  When $q^2 + 4 p^3 < 0$ all three roots are imaginary.  In that case it is better to express the roots somewhat differently.  They are
\ba
{\rm w}_0 &=& -i \frac{\sqrt{3}}{2} \sqrt{-p} \left[ \sqrt{3} \cos(\phi/3) - \sin(\phi/3) \right] \nonumber \\
{\rm w}_1 &=& \oneth x + \thalf \sqrt{-p} \left[ \cos(\phi/3) +\sqrt{3} \sin(\phi/3) \right] \nonumber \\
{\rm w}_3 &=& \oneth x - \sqrt{-p} \left[  \cos(\phi/3) +\sqrt{3} \sin(\phi/3) \right]
\ea
where
\be
\phi = \arccos \left[ \frac{-q}{2(-p)^{3/2}} \right]
\ee
Equivalently, the three roots are
\ba
{\rm w}_+ &=& \oneth x +2\sqrt{-p} \cos(\phi/3) \nonumber \\
{\rm w}_- &=& \oneth x + \sqrt{-p} \left[ -\cos(\phi/3) +\sqrt{3} \sin(\phi/3) \right] \nonumber \\
{\rm w}_3 &=& \oneth x - \sqrt{-p} \left[ \cos(\phi/3) +\sqrt{3} \sin(\phi/3) \right]
\ea

When $k^2 \rightarrow 0$ it is easiest to find the roots from the original cubic equation rather than from the general solutions given above.  In that limit, with $x<2$,
\be
{\rm w}_0 \rightarrow \sqrt{1-x^2/4}
\label{wzero}
\ee
\be
{\rm w}_1 \rightarrow \frac{x}{2}
\ee
\be
{\rm w}_3 \rightarrow \tau_2 D k^2
\ee
The ${\rm w}_3$ represents the long-time diffusion mode.  When $x > 2$, ${\rm w}_0$ begins pure imaginary, and therefore so do $\omega_1$ and $\omega_2$.  

When  $k^2 \rightarrow \infty$  one finds that
\be
\frac{{\rm w}_0}{\tau_2} \rightarrow v_0 k + \frac{(4+3\alpha^2 -2 \alpha x - x^2)}{8 v_0 \tau_2^2 k}
\ee
where $v_0 = \sqrt{\tau_3' D/\tau_2^2}$ and $\alpha = z/y = \tau_2/\tau_3'$.  Furthermore
\ba
{\rm w}_1 &\rightarrow& \frac{x-\alpha}{2} + \frac{\alpha (\alpha^2 - \alpha x +1)}{2v_0^2 \tau_2^2 k^2} \nonumber \\
{\rm w}_3 &\rightarrow& \alpha - \frac{\alpha (\alpha^2 - \alpha x +1)}{v_0^2 \tau_2^2 k^2}
\ea 
Since the poles should lie in the lower half-plane we must insist that $\alpha < x$ or $\tau_2^2 < \tau_1 \tau_3'$. 

Given these limiting behaviors in $k$, it is clear that when $x > 2$ there is a critical value $k_c$ such that for $k < k_c$ the $\omega_1$ and $\omega_2$ are pure imaginary, and when $k > k_c$ they are complex with nonzero real and imaginary parts.

\subsection{Baryon correlation function}

The autocorrelation function gets contributions from all three poles.  In time and wavenumber space, and assuming that ${\rm w}_0$ is real, it can be found to be
\ba
\langle \delta n \delta n (t,{\bf k}) \rangle &=& \frac{\sigma T \tau_2 k^2}{ ({\rm w}_1-{\rm w}_3)^2 + {\rm w}_0^2} \left\{ \left( \frac{1}{{\rm w}_3} - \frac{\tau_4}{\tau_2} \right) 
{\rm e}^{-{\rm w}_3 t/\tau_2} \right. \nonumber \\
&+& \left. \left[ \left( \frac{{\rm w}_1({\rm w}_3-{\rm w}_1) +{\rm w}_0^2}{{\rm w}_1^2 + {\rm w}_0^2} + ({\rm w}_1-{\rm w}_3) \frac{\tau_4}{\tau_2} \right) \frac{\sin({\rm w}_0 t/\tau_2)}{{\rm w}_0} \right. \right. \nonumber \\
&+& \left. \left. \left( \frac{{\rm w}_3 - 2 {\rm w}_1}{{\rm w}_1^2 + {\rm w}_0^2} + \frac{\tau_4}{\tau_2} \right) \cos({\rm w}_0 t/\tau_2) \right] {\rm e}^{-{\rm w}_1 t/\tau_2} \right\}
\label{realw}
\ea
When ${\rm w}_0$ is pure imaginary it can be found to be
\ba
\langle \delta n \delta n (t,{\bf k}) \rangle &=& \frac{\sigma T \tau_2 k^2}{ ({\rm w}_1-{\rm w}_3)^2 - |{\rm w}_0|^2} \left\{ \left( \frac{1}{{\rm w}_3} - \frac{\tau_4}{\tau_2} \right) 
{\rm e}^{-{\rm w}_3 t/\tau_2} \right. \nonumber \\
&+& \left. \left[ \left( \frac{{\rm w}_1({\rm w}_3-{\rm w}_1) -|{\rm w}_0|^2}{{\rm w}_1^2 - |{\rm w}_0|^2} + ({\rm w}_1-{\rm w}_3) \frac{\tau_4}{\tau_2} \right) \frac{\sinh(|{\rm w}_0| t/\tau_2)}{|{\rm w}_0|} \right. \right. \nonumber \\
&+& \left. \left. \left( \frac{{\rm w}_3 - 2 {\rm w}_1}{{\rm w}_1^2 - |{\rm w}_0|^2} + \frac{\tau_4}{\tau_2} \right) \cosh(|{\rm w}_0| t/\tau_2) \right] {\rm e}^{-{\rm w}_1 t/\tau_2} \right\}
\label{imaginaryw}
\ea
Obviously one can go from (\ref{realw}) to (\ref{imaginaryw}) by making the substitution ${\rm w}_0 \rightarrow i |{\rm w}_0|$ in the former.

The conventional diffusion result is recovered in the limit that $r$ is held fixed and $t$ becomes large compared to the characteristic time scales.  Looking back at eqs. (\ref{realw}) and (\ref{imaginaryw}) it is apparent that in this limit the autocorrelation function is dominated by the term involving $\exp(- {\rm w}_3 t /\tau_2)$ as the others are exponentially smaller.  The ${\rm w}_3$ is minimized when $k \rightarrow 0$.  In this limit $({\rm w}_1-{\rm w}_3)^2 + {\rm w}_0^2 =1$ if $x < 2$ and $({\rm w}_1-{\rm w}_3)^2 - |{\rm w}_0|^2 = 1$ if $x > 2$.  The term $\tau_4/\tau_2$ can be ignored in comparison to $1/{\rm w}_3$.  This then leads to eq. (\ref{Dkt}).

Just as in the Cattaneo equation the autocorrelator is an even function of $k$, and has similar large $k$ behavior.  The dispersion relation at large $k$ is ${\rm w}_0 = \sqrt{\tau_3' D}k$ so that when $r > v_0 t$ the autocorrelation function vanishes.  There is a step function and a Dirac delta function and its derivatives located at $r=v_0 t$, followed by a wake.  To show some numerical results we shall take $\tau_4=0$ to shorten the formulas.  We use dimensionless variables ${\hat r}=r/\tau_2$, ${\hat k}=\tau_2 k$, and ${\hat t}=v_0 t/\tau_2$.  Then $y=v_0^2 {\hat k}^2$ and $z = \alpha v_0^2 {\hat k}^2$.  Equation (\ref{realw}) becomes
\be
\langle \delta n \delta n (t,{\bf k}) \rangle = \frac{\sigma T}{2 \pi^2 v_0^2 \tau_2^4}\, f({\hat r},{\hat t})
\ee
where
\ba
 f({\hat r},{\hat t}) &=& \frac{1}{{\hat r}} \int_0^{\infty} d{\hat k} \, {\hat k} \, \sin({\hat k}{\hat r}) \,
\frac{v_0^2 {\hat k}^2}{ ({\rm w}_1-{\rm w}_3)^2 + {\rm w}_0^2}  \nonumber \\
&\times& \left\{ \frac{{\rm e}^{-{\rm w}_3 {\hat t}/v_0}}{{\rm w}_3} 
+ \left[ \left( \frac{{\rm w}_1({\rm w}_3-{\rm w}_1) +{\rm w}_0^2}{{\rm w}_0({\rm w}_1^2 + {\rm w}_0^2)}  \right) \sin({\rm w}_0 {\hat t}/v_0) \right. \right. \nonumber \\
&+& \left. \left. \left( \frac{{\rm w}_3 - 2 {\rm w}_1}{{\rm w}_1^2 + {\rm w}_0^2}  \right) \cos({\rm w}_0 {\hat t}/v_0) \right] {\rm e}^{-{\rm w}_1 {\hat t}/v_0} \right\}
\label{realwf}
\ea
When ${\rm w}_0$ is imaginary the integrand changes to the form given in eq. (\ref{imaginaryw}).  From the integrand must be subtracted the large ${\hat k}$ limit in order to obtain an integral that can be done numerically.  Apart from the overall factor ${\hat k} \sin({\hat k}{\hat r})$ what must be subtracted is
\bd
\left[ 1 + \frac{\alpha (\alpha^2 - \alpha x +1)}{v_0^3 {\hat k}^2}  {\hat t} + \frac{\alpha(x-2\alpha)}{v_0^2 {\hat k}^2} \right] 
\frac{{\rm e}^{-\alpha {\hat t}/v_0}}{\alpha}
\ed
\bd
+ \left[ \frac{\sin({\hat k}{\hat t})}{v_0 {\hat k}}+\left[ (2\alpha - x)v_0 + {\textstyle{\frac{1}{8}}}
(4 + 3\alpha^2 - 2 \alpha x - x^2) {\hat t} \right]
\frac{\cos({\hat k}{\hat t})}{v_0^3 {\hat k}^2} \right] 
{\rm e}^{-(x-\alpha) {\hat t}/2v_0}
\ed
The first term in square brackets in the first line leads to a singular term proportional to $\delta ({\hat {\bf x}}) {\rm e}^{-\alpha {\hat t}/v_0}$.  The first term in square brackets in the second line leads to a singular term proportional to $ {\hat r}^{-1} \delta ({\hat r}-{\hat t}) {\rm e}^{-(x-\alpha) {\hat t}/2v_0}$.  The remaining terms are finite and should be added back to obtain the regular part of $f$.  They are
\bd
\frac{\pi}{2 v_0^3 } \left[ (\alpha^2 - \alpha x +1) {\hat t} + v_0 (x -2 \alpha) \right] 
\frac{{\rm e}^{-\alpha {\hat t}/v_0}}{{\hat r}}
\ed
and 
\bd
\frac{\pi}{16 v_0^3 } \left[8 (2\alpha - x)v_0 + 
(4 + 3\alpha^2 - 2 \alpha x - x^2) {\hat t} \right]
\frac{{\rm e}^{-(x-\alpha) {\hat t}/2v_0}}{{\hat r}} \theta ({\hat r}-{\hat t})
\ed

\begin{figure}
  \centering
  \includegraphics{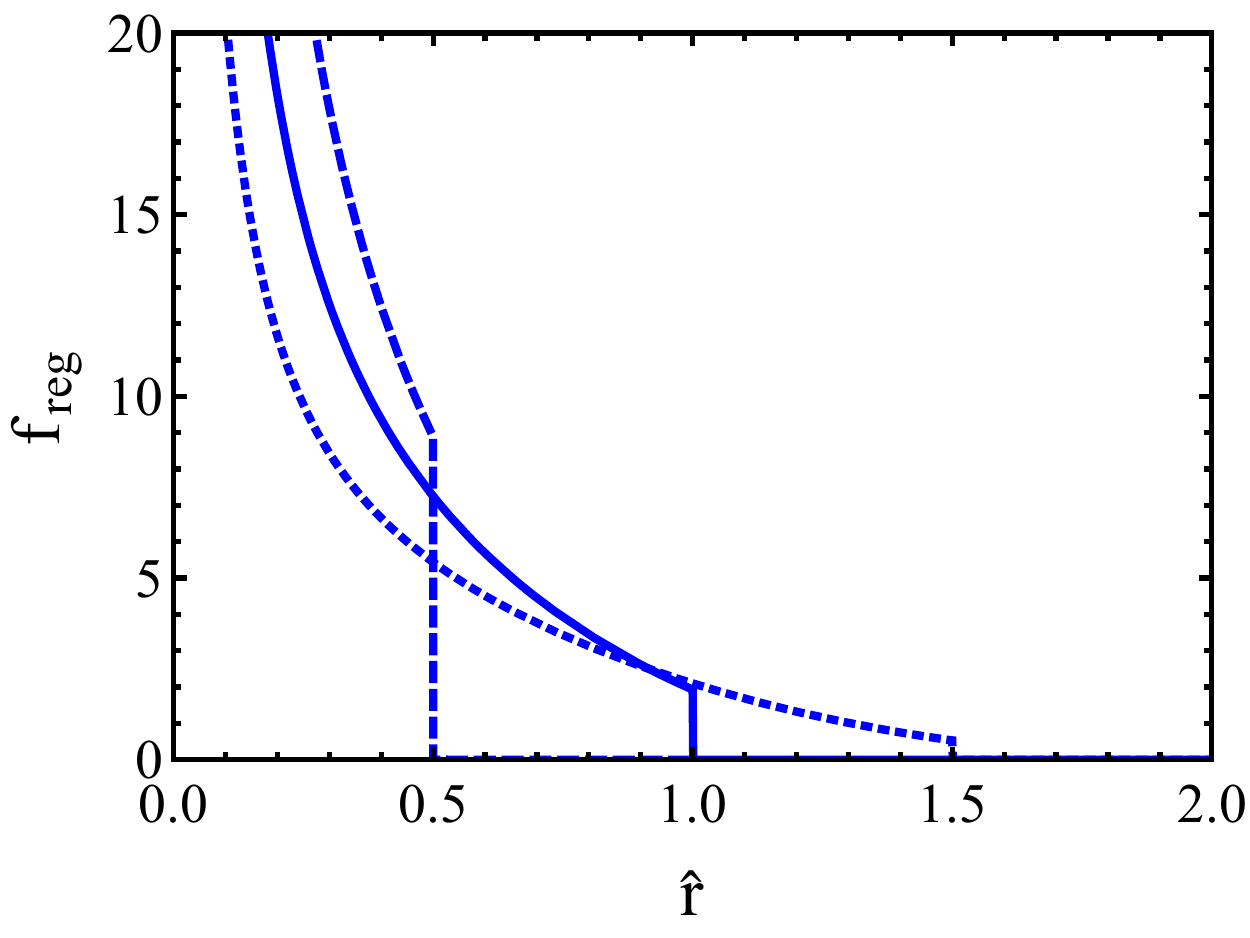}
  \caption{The regular part of the dimensionless baryon correlator $f_{\rm reg}(\hat{r},\hat{t})$ for the Gurtin-Pipkin equation with $x=3$.
  The correlator is shown for $\hat{t}=0.5$ (dashed line),  $\hat{t}=1$ (solid line), and  $\hat{t}=1.5$ (dotted line).}
  \label{fig:nn_GP_JK8}
\end{figure}

\begin{figure}
  \centering
  \includegraphics{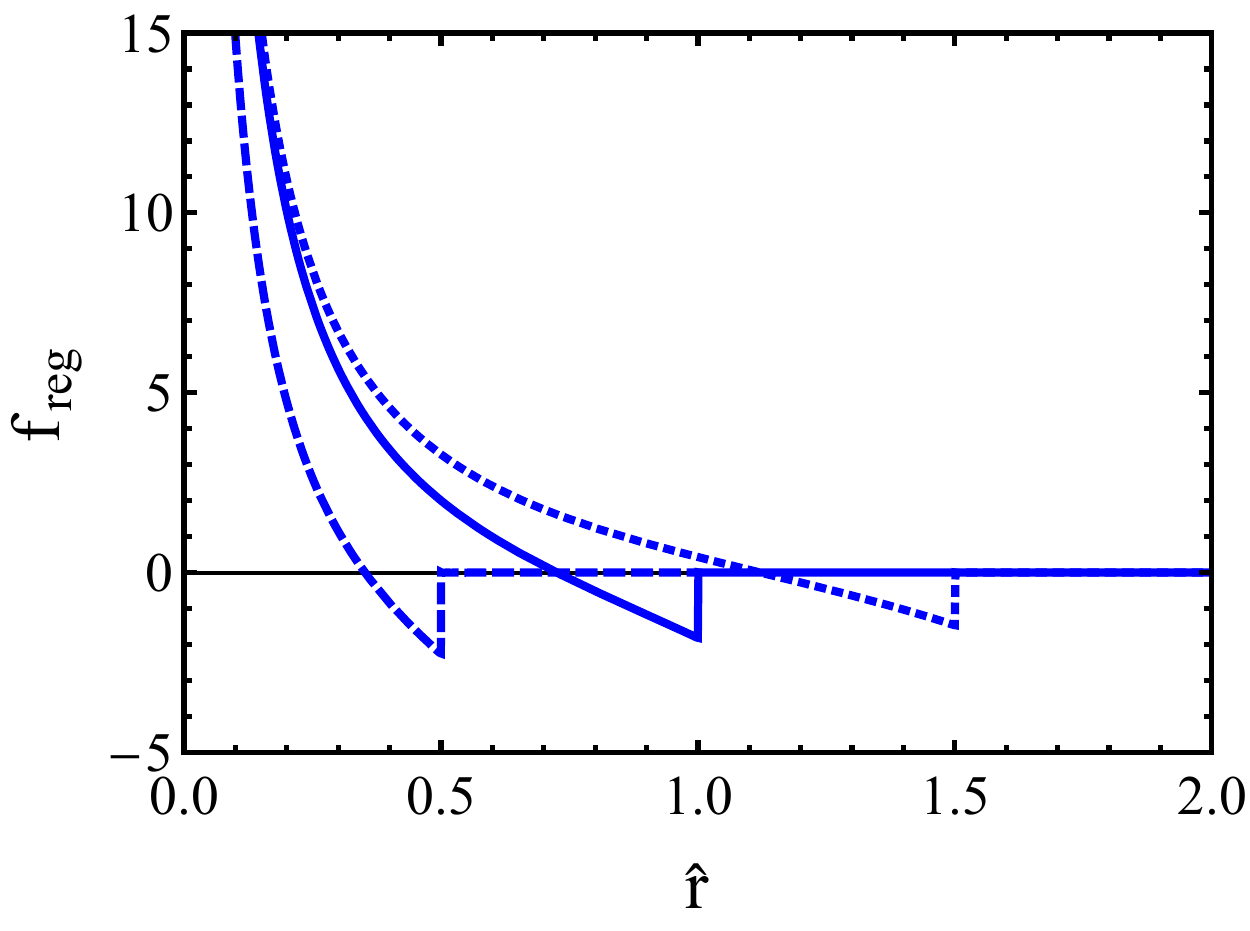}
  \caption{The regular part of the dimensionless baryon correlator $f_{\rm reg}(\hat{r},\hat{t})$ for the Gurtin-Pipkin equation with $x=1$.
  The correlator is shown for $\hat{t}=0.5$ (dashed line),  $\hat{t}=1$ (solid line), and  $\hat{t}=1.5$ (dotted line).}
  \label{fig:nn_GP_JK7}
\end{figure}

For illustration we show $f_{\rm reg}(\hat{r},\hat{t})$ as a function of $\hat{r}$ for three different times in figures \ref{fig:nn_GP_JK8} (x=3) and \ref{fig:nn_GP_JK7} (x=1).  In both cases we chose $v_0^2 = 1/3$.  They show the characteristic wake following the front located at ${\hat r}={\hat t}$.


\subsection{Noise}

In frequency and wavenumber space the noise correlator is obtained from eqs. (\ref{gen_noise_cor}) and (\ref{A}).  The function $A(\omega,k)$ has poles at
\ba
\omega_{\pm} &=& -\frac{i \tau_1}{2 \tau_2^2} \pm \frac{i \delta_k}{2\tau_2} \nonumber \\
\delta_k &=& \sqrt{\frac{\tau_1^2}{\tau_2^2} - 4 \left( 1 + \tau_3 D k^2 \right)}
\label{largex}
\ea
when $k^2 < k_c^2 \equiv (\tau_1^2/\tau_2^2 - 4)/4 \tau_3 D$.  When $k^2 > k_c^2$ the poles are at
\ba
\omega_{\pm} &=& -\frac{i \tau_1}{2 \tau_2^2} \pm \frac{\epsilon_k}{2\tau_2} \nonumber \\
\epsilon_k &=& \sqrt{4 \left( 1 + \tau_3 D k^2\right) - \frac{\tau_1^2}{\tau_2^2}}
\label{smallx}
\ea
When $x < 2$ the poles are always complex and only (\ref{smallx}) applies, not (\ref{largex}).  This is the same situation as for the baryon autocorrelator.  

The group velocity
\be
v_g = \frac{2(\tau_3/\tau_2)Dk}{\sqrt{4 \left( 1 + \tau_3 D k^2\right) - x^2}}
\ee
is defined only for $k > k_c$ if $x > 2$.  For $x < 2$ it is defined for all $k$.  For $x > 2$ it diverges at $k_c$ which is the same situation as in the autocorrelator in the Cattaneo equation.  For $x<2$ the group velocity is always below its asymptotic value of 
$v_0 = \sqrt{\tau_3 D/\tau_2^2}$.  Note that this differs with the asymptotic speed in the baryon autocorrelator if $\tau_4 \neq 0$.

In time ($t>0$) and wavenumber space the noise correlator is
\be
\oneth \langle I^l I^l ({\bf k},t) \rangle = \frac{\sigma T}{\tau_2}
\left[ \frac{\tau_4}{\tau_2} \cosh\left(\frac{\delta_k t}{2\tau_2}\right)
+ \left(2-\frac{\tau_1 \tau_4}{\tau_2^2}\right) \frac{1}{\delta_k}\sinh\left(\frac{\delta_k t}{2\tau_2}\right) \right]
\exp\left(-\frac{\tau_1 t}{2\tau_2^2}\right)
\ee
when $k<k_c$ and
\be
\oneth \langle I^l I^l ({\bf k},t) \rangle = \frac{\sigma T}{\tau_2}
\left[ \frac{\tau_4}{\tau_2} \cos\left(\frac{\epsilon_k t}{2\tau_2}\right)
+ \left(2-\frac{\tau_1 \tau_4}{\tau_2^2}\right) \frac{1}{\epsilon_k}\sin\left(\frac{\epsilon_k t}{2\tau_2}\right) \right]
\exp\left(-\frac{\tau_1 t}{2\tau_2^2}\right)
\ee
when $k>k_c$.  These have some interesting analytical limits.

Setting $\tau_4 = 0$, letting $\tau_2 \rightarrow 0$, and then setting $\tau_3 = 0$ (so that the condition $\tau_2^2 < \tau_1 \tau_3'$ is respected), one obtains the Cattaneo result eq. (\ref{Cnoise}).  

Setting $\tau_1 = 2\tau_2$, equivalently $x=2$, results in $\epsilon_k = 2 v_0 \tau_2 k$.  Then
\be
\oneth \langle I^l I^l ({\bf k},t) \rangle = \frac{\sigma T}{\tau_2}
\left[ \frac{\tau_4}{\tau_2} \cos\left(v_0 k t\right)
+ \left(1-\frac{\tau_4}{\tau_2}\right) \frac{\sin\left(v_0 k t\right)}{v_0 \tau_2 k} \right]
\exp\left(-\frac{ t}{\tau_2}\right)
\ee
whose Fourier transform is
\be
\oneth \langle I^l I^l ({\bf x},t) \rangle = \frac{\sigma T}{4 \pi v_0 \tau_2^2 r} \left[ \left(1-\frac{\tau_4}{\tau_2}\right) \delta (r - v_0 t) - v_0 \tau_4 \delta' (r - v_0 t) \right]
\ee
This is just a pulse which, interestingly, leaves behind no wake.

\begin{figure}
  \centering
  \includegraphics{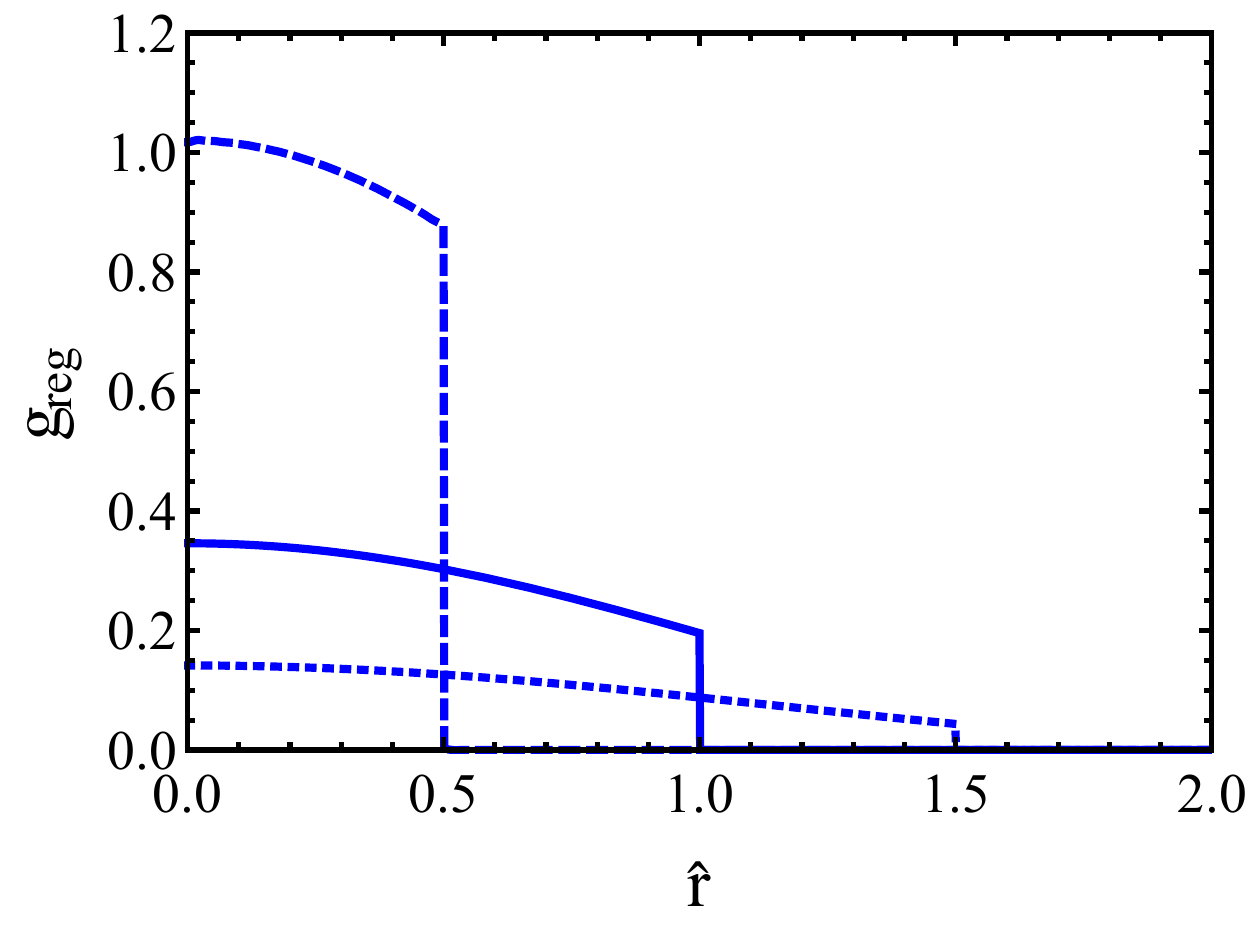}
  \caption{The regular part of the dimensionless noise correlator $g_{\rm reg}(\hat{r},\hat{t})$ for $x=3$.
  The correlator is shown for $\hat{t}=0.5$ (dashed line),  $\hat{t}=1$ (solid line), and  $\hat{t}=1.5$ (dotted line).}
\label{fig:II_x=3}
\end{figure}

\begin{figure}
  \centering
  \includegraphics{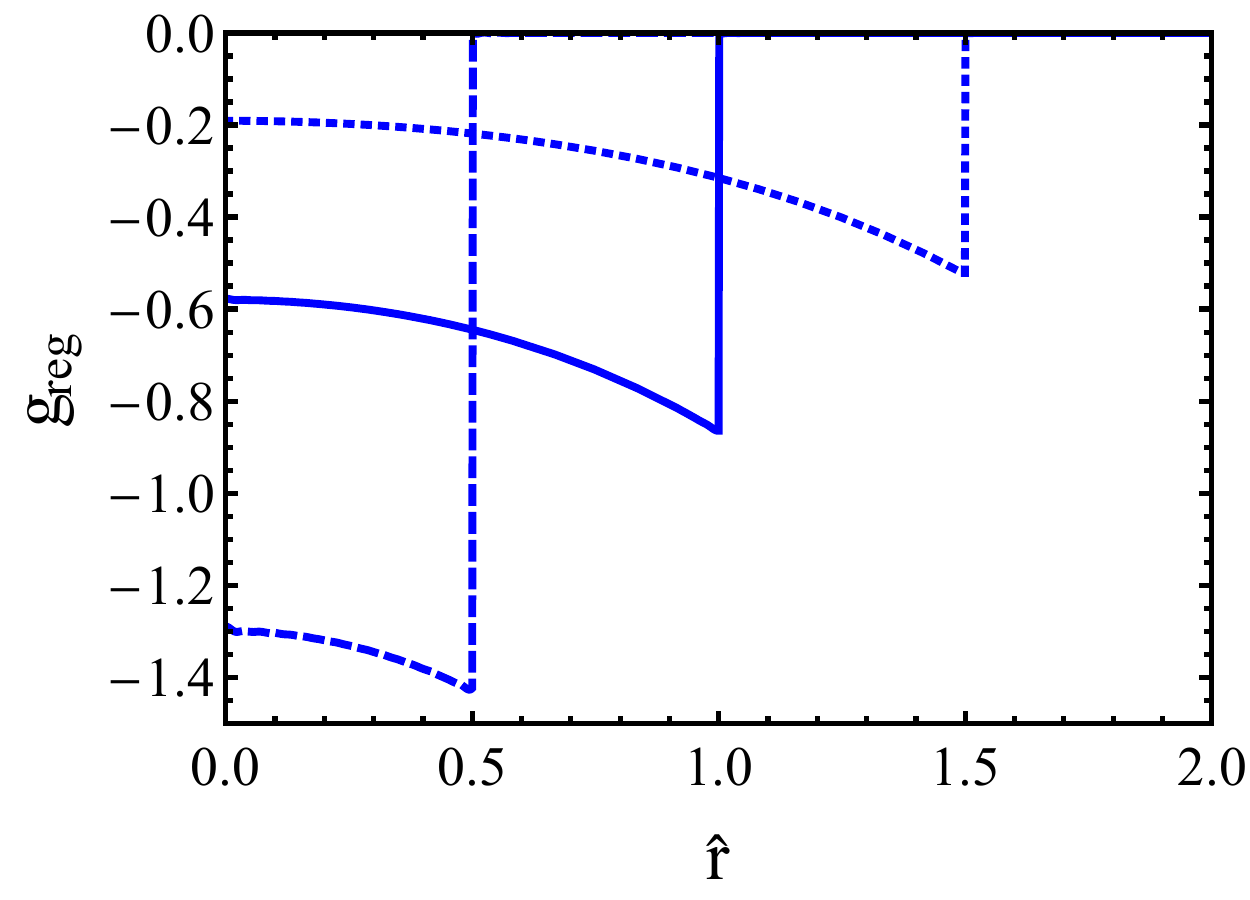}
  \caption{The regular part of the dimensionless noise correlator $g_{\rm reg}(\hat{r},\hat{t})$ for $x=1$.
  The correlator is shown for $\hat{t}=0.5$ (dashed line),  $\hat{t}=1$ (solid line), and  $\hat{t}=1.5$ (dotted line).}
\label{fig:II_x=1}
\end{figure}

From now on we shall consider only the situation $\tau_4 = 0$.  This simplifies presentation of the main features, and is rather natural since it leads to the same speed of propagation of noise as baryon density fluctuations, although in principle they need not be the same.  To display numerical results it is convenient to use the dimensionless variables ${\hat t} = t/2\tau_2$, ${\hat k} = 2 v_0 \tau_2 k$, and ${\hat r} = r/2 v_0 \tau_2$.  Then
\be
\oneth \langle I^l I^l ({\bf x},t) \rangle = \frac{\sigma T}{8 \pi^2 v_0^3 \tau_2^4} \, g({\hat r},{\hat t})
\ee
where
\be
g(\hat{r},\hat{t}) = \frac{{\rm e}^{-x\hat{t}}}{\hat{r}} \int_0^{\infty} d\hat{k} \, \hat{k} \sin(\hat{k}\hat{r}) \frac{\sin(\epsilon_{\hat k} {\hat t})}{\epsilon_{\hat k}}
\ee
Hidden within the integral are singularities at the point ${\hat r} = {\hat t}$.  These may be extracted by examining the large ${\hat k}$ limit.  
\be
\frac{\sin(\epsilon_{\hat k} {\hat t})}{\epsilon_{\hat k}} \rightarrow \frac{\sin({\hat k} {\hat t})}{{\hat k}} + 
\frac{(4-x^2){\hat t}}{2{\hat k}^2} \cos({\hat k} {\hat t})
\ee
These  terms contribute to $g$ as
\bd
\frac{\pi {\rm e}^{-x\hat{t}}}{4 {\hat r}}  \left[ 2 \delta ({\hat r}-{\hat t}) + (4-x^2) \,{\hat t}\, \theta  ({\hat r}-{\hat t}) \right]
\ed
The regular part of $g$, which does not include the Dirac delta function but which does include the step function, is
\ba
g_{\rm reg}(\hat{r},\hat{t}) &=& \frac{{\rm e}^{-x\hat{t}}}{\hat{r}} \left\{ 
\int_0^{\infty} d\hat{k} \, \hat{k} \sin(\hat{k}\hat{r}) \left[ \frac{\sin(\epsilon_{\hat k} {\hat t})}{\epsilon_{\hat k}}
- \frac{\sin({\hat k} {\hat t})}{{\hat k}} - \frac{(4-x^2){\hat t}}{2{\hat k}^2} \cos({\hat k} {\hat t})  \right] \right. \nonumber \\
&+& \left. \frac{\pi (4-x^2) {\hat t}}{4} \theta  ({\hat r}-{\hat t})  \right\}
\ea
Some representative plots of $g_{\rm reg}$ are shown in Figures \ref{fig:II_x=3} and \ref{fig:II_x=1}.  For $x>2$ the wake behind the front represents a positive correlation, while for $x<2$ it represents a negative correlation; for $x=2$ the wake is absent, as mentioned above.

\section{Conclusions}
\label{conclusions}

In this paper we studied and compared the baryon current in first, second, and third order dissipative fluid dynamics using the Landau-Lifshitz definition of flow velocity.  With no energy transport but only pure baryon diffusion, the resulting equations correspond to the ordinary heat conduction equation, the Cattaneo heat conduction equation, and the Gurtin-Pipkin heat conduction equation, respectively.   Using the fluctuation-dissipation theorem we computed the response function, the baryon autocorrelation function, and the correlation for thermal noise.  Unlike the case of the first order theory, the second and third order theories progagate signals with a finite speed.  The parameters in the theories must be such that the group velocity at large wavenumbers does not exceed the speed of light, for it is this velocity with which signals travel.

Previous work examined the effect of diffusion beyond first order on correlations in heavy-ion collisions.  Reference \cite{Aziz} found that using the Cattaneo equation instead of the first order diffusion equation lessens the extent to which diffusion can dissipate fluctuations. Enforcing causality makes clear the importance of fluctuations at early times for creating long-range correlations in rapidity.  The thermal noise discussed here is one source for these fluctuations. We have also demonstrated how including the additional transport coefficients of the Gurtin-Pipkin equation might also have an effect on correlation functions.

In the first order theory the noise correlator is proportional to a product of Dirac delta functions in space and time, as has been known for a long time.  For hydrodynamic modeling of high energy heavy ion collisions this would generally be sufficient if noise is treated as a perturbation.  However, in some situations, such as near a critical point, noise may play such an important role that it should be treated nonperturbatively in the hydrodynamic evolution.  In that case, it is important to have a noise correlator that is of finite range in time and space lest the results become sensitive to the size of the coarse-graining cells.  The second order theory has a finite range in time but is still a delta function in space.  One must go to the third order to have a finite range in space as well.

We have not attempted to deduce the $\tau$ parameters in the third order theory.  In principle they should be calculable from a microscopic theory.  Undoubtedly they will be functions of temperature and density.  Of course one should expect coupling of the current to various components of the energy-momentum tensor when the temperature and flow velocity can vary in space and time; this might provide relationships between these $\tau$ parameters and others that appear in higher order viscous fluid dynamics \cite{Israel,Denicol2012}.  We look forward to future progress in these and other avenues of investigation.   

\acknowledgments

We thank K. Dusling for discussions.  This work was supported by the U.S. DOE Grant No. DE-FG02-87ER40328.

\end{document}